\documentclass[a4paper,11pt]{article}
\usepackage{jheppub} 
\usepackage{tikz-cd}

\title{Quantum harmonic oscillator, index theorem and spectral asymmetry}

\author[1]{Shunrui Li} 
\author[2,3]{and Yang Liu}

\affiliation[1]{Department of mathematics, Free University of Berlin, Berlin 14195, Germany}
\affiliation[2]{School of Physics and Astronomy, University of Nottingham, University Park, Nottingham NG7 2RD, United Kingdom}
\affiliation[3]{Department of Physics, Tsinghua University, Beijing 100084, China}

\emailAdd{lis02@zedat.fu-berlin.de}
\emailAdd{liu-yang\_1990@mail.tsinghua.edu.cn}

\abstract{
We report a spectral asymmetry effect in the quantum harmonic oscillator, where its partition function is identified as the Chern character. This establishes a direct link between statistical mechanics, and topological invariants (Atiyah-Singer index theorem), revealing the internal energy as a non-SUSY manifestation of the index theorem. We show that the partition function can be interpreted as the Chern character of “virtual physical sheaf”, namely, a Hermitian vector bundle encoding quantum states over spacetime. This work uncovers an underlying topological structure in bosonic quantum systems.}

\begin{document}
\maketitle
\flushbottom
\section{Introduction}

The Atiyah-Singer index theorem \cite{atiyah2006elliptic, atiyah1963index, ABP1973} and its generalizations have profoundly shaped modern theoretical physics, establishing deep connections between the analytical properties of differential operators and the topological invariants of underlying manifolds. In quantum field theory (QFT) and condensed matter physics, these topological frameworks have been instrumental in explaining physical phenomena such as chiral anomalies, topological insulators, and the fractional quantum Hall effect. However, the application of index theorems in physics has traditionally been dominated by fermionic systems and supersymmetric (SUSY) theories, where the Dirac operator naturally maps to topological characteristic classes. The intrinsic topological structure of purely bosonic systems, particularly within the context of finite-temperature statistical thermodynamics, remains a comparatively less explored frontier \cite{nakahara2018geometry}.

In this paper, we investigate the foundational quantum harmonic oscillator (QHO) to reveal an unexpected, exact algebraic correspondence between bosonic statistical mechanics and topological index theory. Rather than relying on supersymmetry or fermionic zero modes, we demonstrate that the thermal partition function and the internal energy of the QHO fundamentally encode the generating functions of topological characteristic classes, specifically the Hirzebruch $\mathcal{L}$-genus and the $\hat{A}$-genus.

To establish this correspondence rigorously without falling into the functional-analytic pitfalls of infinite-dimensional geometric topologies, we adopt a formal algebraic approach rooted in geometric quantization. When a quantum system is coupled to a thermal bath, Euclidean time is compactified into a thermal circle $S^1_\beta$. In this setting, the Hamiltonian acts as the curvature of a temporal connection. We abstract the infinite-dimensional quantum state space into a formal algebraic structure, which we term the \textit{virtual physical sheaf}. By endowing this virtual sheaf with the functorial properties of K-theory, the QFT thermal trace over Matsubara frequencies algebraically mirrors the geometric evaluation of the Chern character. 


The paper is organized as follows: In Section 2, we briefly review the generalized Hirzebruch signature theorem. In Section 3, we establish the algebraic connection between the dimensionless internal energy of the QHO and the $\mathcal{L}$-genus. In Section 4, we construct the virtual physical sheaf, resolve the finite-to-infinite dimensional topological limit, and use the GRR theorem to identify the partition function as Chern characteristic. Finally, in Section 5, we analyze the non-invariance of the topological index under orientation reversal, formally introducing the Topological Spectral Asymmetry Effect.

\section{The generalized Hirzebruch signature theorem}
In this paper, we explore the connection between the quantum harmonic oscillator and a specific case of the Atiyah–Singer Index Theorem, known as the generalized Hirzebruch signature theorem. We begin with a brief overview of this theorem.
Let $M$ be a compact, oriented Riemannian manifold of dimension $2l$, and let $\xi$ be a Hermitian vector bundle over $M$. Let $A_{\xi}$ represent the generalized signature operator corresponding to $\xi$. The index of $A_{\xi}$ is given by the following expression:
\begin{equation} \label{GHST}
\text{index} \ A_{\xi} = 2^l \cdot \text{ch} \xi \cdot \mathcal{L} (M) [M],
\end{equation}
where $\text{ch} \xi$ is the Chern character of the bundle $\xi$, and $\mathcal{L} (M)$ is the Hirzebruch $\mathcal{L}$-polynomial, defined by
\begin{equation} \label{LM}
\mathcal{L} (M) = \prod^l_{i=1} \frac{x_i/2}{\tanh{x_i /2}}.
\end{equation}
The elementary symmetric functions of the $x^2_i$ are replaced by the Pontrjagin classes of $M$ \cite{ABP1973}.
Here, the $x_i$ are formal variables whose elementary symmetric functions in $x^2_i$ correspond to the Pontryagin classes of the manifold $M$ \cite{ABP1973}.

\section{An example: the internal energy of quantum harmonic oscillator}
Let us begin by revisiting a fundamental example in physics: the quantum harmonic oscillator. The energy levels of a quantum harmonic oscillator are given by: 
\begin{equation} \label{En}
E_n=(n+\frac{1}{2})\hbar \omega,
\end{equation}
where $\omega$ is angular frequency of oscillator. The partition function of this system is computed as:
\begin{equation} \label{Z1}
Z = \sum^{\infty}_{n=0} e^{-\beta E_n} = \sum^{\infty}_{n=0} e^{-\beta (n + \frac{1}{2}) \hbar \omega}=\frac{e^{\frac{-\beta \hbar \omega}{2}}}{1 - e^{-\beta \hbar \omega}} = \frac{1}{ e^{\frac{1}{2} \beta \hbar \omega} - e^{-\frac{1}{2} \beta \hbar \omega}} 
\end{equation}
where $\beta = \frac{1}{k T}$. The total energy of the system is then given by the standard thermodynamic relation:
\begin{equation} \label{U}
U \equiv - \frac{\partial}{\partial \beta} \ln Z = - \frac{1}{Z} \frac{\partial}{\partial \beta}  Z,
\end{equation}
namely,
\begin{equation} \label{U0}
U_0=\frac{\varepsilon}{2}+\frac{\varepsilon}{e^{\varepsilon / k T}-1}, \quad \text{where} \quad \epsilon=\hbar \omega.
\end{equation}
To reveal the underlying topological structure of the system, let us examine its natural dimensionless thermodynamic quantities. We define the dimensionless parameter $x = \beta \epsilon = \beta \hbar \omega$, which represents the ratio of the energy quantum to the thermal energy. The internal energy $U$ can then be naturally rescaled into a dimensionless form, $\beta U$. By multiplying the standard internal energy relation by $\beta$, we obtain:
\begin{equation}\label{beta U}
    \beta U = \frac{\beta \epsilon}{e^{\beta \epsilon}-1} + \frac{\beta \epsilon}{2} = \frac{x}{e^x - 1} + \frac{x}{2} = \frac{x}{2} \frac{e^{x / 2}+e^{-x / 2}}{e^{x / 2}-e^{-x / 2}} = \frac{x/2}{\tanh(x/2)}.
\end{equation}
This final expression is precisely the generating function of the topological invariant $\mathcal{L}$-genus, $\mathcal{L}(x)$. Crucially, this equivalence emerges naturally for any arbitrary temperature $\beta$, demonstrating that the topological structure is a robust, universal scaling property of the system rather than an artifact of choosing specific units. This final expression intriguingly echoes the form of characteristic classes appearing in index theory, suggesting deep connections between statistical thermodynamics and topological invariants.

In fact, the calculation presented in Section 3 can be naturally connected to the generalized Hirzebruch signature theorem. Let us now clarify this connection. We can notice a fact that the definitions of the Chern character, given by
\begin{equation} \label{chbetaxi}
    \text{ch} (\beta \xi) = \sum_i e^{\beta x_i},
\end{equation}
and the partition function
\begin{equation} \label{Z}
    Z = \sum_n e^{-\beta E_n},
\end{equation}
share a striking formal similarity. 
By identifying the scaled energy eigenvalues $-\beta E_n$ with the formal roots $x_i$ (or equivalently, identifying the scaled Hamiltonian $-\beta H$ with the curvature form), the partition function inherently computes the Chern character of the corresponding state space. Consequently, the structures of the thermodynamic derivation and the index formula are fundamentally isomorphic. 

\subsection{Internal energy $U$, $\hat{A}$-genus and $\mathcal{L}$-genus}
To further explore the algebraic correspondence between the thermodynamic quantities of the QHO and topological characteristic classes, we establish a formal connection between the generating functions of the $\mathcal{L}$-genus and the $\hat{A}$-genus (A-roof genus).

Let us define the dimensionless parameter $x = \beta \hbar \omega$. The generating function for the $\hat{A}$-genus is defined as:
\begin{equation} \label{Ahat}
\hat{A}(x) = \frac{x/2}{\sinh{(x/2)}}.
\end{equation}
Remarkably, the thermal partition function of the quantum harmonic oscillator can be formally rewritten in terms of this $\hat{A}$-genus generating function:
\begin{equation} \label{partitionQHO}
Z = \frac{1}{2\sinh(x/2)} = \frac{\hat{A}(x)}{x}.
\end{equation}
Using this structural form, we compute the dimensionless internal energy $\beta U = - \beta \frac{\partial \ln Z}{\partial \beta} = - x \frac{\partial \ln Z}{\partial x}$. Substituting equation \eqref{partitionQHO} into the thermodynamic definition, we obtain:
\begin{equation} \label{internalenergyQHO}
\beta U = \frac{x}{2} \frac{e^{x/2} + e^{-x/2}}{e^{x/2} - e^{-x/2}} = \frac{x/2}{\tanh(x/2)}.
\end{equation}
By factoring this expression, we reveal a beautiful algebraic relation:
\begin{equation} \label{UandL}
\beta U = \left( \frac{x/2}{\sinh(x/2)} \right) \cosh(x/2) = \hat{A}(x) \cosh(x/2) \equiv \mathcal{L}(x).
\end{equation}
Thus, without assuming any specific temperature (i.e., for any arbitrary $\beta$), $\beta U$ strictly equals the $\mathcal{L}$-genus generating function $\mathcal{L}(x)$. Furthermore, equation \eqref{UandL} perfectly reproduces the well-known topological identity relating the $\mathcal{L}$-genus to the $\hat{A}$-genus twisted by a formal hyperbolic cosine factor. While the $\hat{A}$-genus is typically associated with the index of Dirac operators for fermions, its formal appearance here in the bosonic partition function highlights a profound algebraic universality crossing different quantum statistics.
\section{Partition function and Chern character}
In this section, we reinterpret the partition function as the Chern character of a virtual sheaf of wave functions over spacetime. Physically, this sheaf organizes wave functions into local
Hilbert spaces. We can define the virtual sheaf as “virtual physical sheaf”, namely, let $\mathcal{S}\to M$ be a “virtual physical sheaf” encoding quantum states, with fibers isomorphic to $L^2$ at each spacetime point. For the harmonic oscillator, sections of $\mathcal S$ are solutions $\psi(x,t)$ of $(i\partial_t-H)\psi=0$ on $\mathbb R\times S^1_\beta$; 
each energy eigenmode $\psi_n(x)e^{-iE_n t}$ is a global section and the eigenmodes span the space of global sections. 
(If time is Euclidean/thermal, replace $e^{-iE_n t}$ by Matsubara modes $e^{i\omega_m\tau}$ with $\omega_m=2\pi m/\beta$ for bosons.)

Considering a solution $\psi(x)$ to the Schrödinger equation, we expand it in terms of a complete orthonormal basis $\psi_i(x)$ of the Hilbert space as follows:
\begin{equation} \label{expansion}
    \psi(x)=\sum_n c_i \psi_i(x),
\end{equation}
where \( c_i = \langle \psi_i(x), \psi(x) \rangle_{L^2} \) are the expansion coefficients, and $\psi(x)$ is normalized wave function. 

Furthermore, this expansion is expected to converge in the \( L^2 \) sense, meaning that:
\begin{equation} \label{L2 convergent}
    \lim_{n \to \infty} \left\| \psi(x) - \sum_{i=0}^{n} c_i \psi_i(x) \right\|_{L^2} = 0.
\end{equation}
We begin by considering the Hamiltonian operator $H$, which corresponds to $A_{\xi}$ in the generalized Hirzebruch signature theorem:
\begin{equation} \label{section H A1}
    H: \Gamma(M, \xi) \rightarrow \Gamma(M, \xi).
\end{equation}
We regard the Hamiltonian operator as a mapping acting on the space composed of quantum states. Geometrically, we can  view Hamiltonian operator as acting on the section of a vector bundle, namely, the Hamiltonian operator acts on the quantum state space by evolving wave functions in time.\footnote{The action of $H$ evaluates the spectral curvature of the virtual sheaf component by component. This explicitly identifies the analytical trace of the Euclidean evolution with the topological index, setting the stage for a K-theoretic geometric quantization.}
We may interpret this action as operating on the sections of a sheaf $\mathcal{S} \rightarrow M$, where $M$ denotes the spacetime manifold with coordinates $(x,t)$. That is:
\begin{equation}\label{section H A2}
    H:\psi(x,t)\in\Gamma(M,\mathcal{S}) \mapsto H\psi(x,t),
\end{equation}
where $M$ denotes the spacetime manifold with coordinates $(x,t)$, and $\xi$ is a vector bundle over $M$. Our next goal is to establish the relationship between the bundle $\xi$ and the solutions of operator $H$.
In this picture, the Schrödinger equation describes how sections of $\mathcal{S}$ evolve under the Hamiltonian flow. Our next step is to clarify the relation between this sheaf $\mathcal{S}$ and the spectral structure of the Hamiltonian.
\subsection{Finite dimensional case}
To truncate the spectrum at the first $n+1$ levels, consider all wave functions of the form
$\displaystyle \psi(x)=\sum_{i=0}^n a_i\,\psi_i(x)$
with complex coefficients $a_i$. All physical states can be viewed as normal vectors in $\mathcal{O}(1)$ over $\mathbb{CP}^n$. On $\mathbb{CP}^n$, the vector bundle $\mathcal{O}(1)$ carries the standard Fubini–Study Hermitian metric and connection.
To explicitly describe the Chern connection of the normal bundle over the complex projective space \(\mathbb{CP}^n\), which can also be viewed as a finite-dimensional complex sphere, we must first provide a concrete representation of the vector bundle on \(\mathbb{CP}^n\) along with its associated Hermitian metric. \\
1. Complex projective space \(\mathbb{CP}^n\) and hyperplane bundles.\\
The complex projective space \(\mathbb{CP}^n\) can be defined as:
\begin{equation} \label{CPn}
\mathbb{CP}^n = \frac{\mathbb{C}^{n+1} \setminus \{0\}}{\mathbb{C}^*},
\end{equation}
where \(\mathbb{C}^{n+1}\) is a complex vector space of \((n+1)\)-dimension, and \(\mathbb{C}^*\) denotes the multiplicative group of non-zero complex numbers. Each point \([z] \in \mathbb{CP}^n\) corresponds to an equivalence class of nonzero vectors under scalar multiplication, known as a homogeneous coordinate. The hyperplane bundle \(\mathcal{O}(1)\) plays a fundamental role on \(\mathbb{CP}^n\), with its dual bundle \(\mathcal{O}(-1)\) forming a subbundle of the trivial bundle $\mathbb{CP}^n \times \mathbb{C}^{n+1}$. \\
2. Definition of normal bundle. \\
Consider the embedding relation of each tangent bundle in complex projective space. For the embedding \(\mathbb{CP}^n \hookrightarrow \mathbb{C}^{n+1}\), we have the following exact sequence:
\begin{equation} \label{embedding CPn to Cn+1}
0 \to T\mathbb{CP}^n \to \mathbb{C}^{n+1} \times \mathbb{CP}^n \to \mathcal{O}(1) \to 0,
\end{equation}
where \(\mathcal{O}(1)\) denotes the hyperplane bundle, and \(\mathbb{C}^{n+1} \times \mathbb{CP}^n\) is the trivial bundle. The normal bundle \(\mathcal{N}\) is the complement bundle that ensures the exactness of the above short exact sequence. Therefore, we have: 
\begin{equation} \label{bundle N}
\mathcal{N} \cong \mathcal{O}(1). 
\end{equation} 

The Hermitian metric on $\mathcal{O}(1)$ naturally defines a Chern connection, whose curvature two-form is proportional to the Fubini–Study Kähler form \cite{nakahara2018geometry}
\begin{equation} \label{chern curvature and FS form}
F = - 2\pi i \omega_{FS}.
\end{equation}

In the framework of geometric quantization \cite{K1970,SJM1970}, the classical phase space is endowed with a symplectic structure, which upon quantization corresponds to the curvature of a prequantum line bundle. When the eigenstates are finite, we can interpret the Hamiltonian operator in the harmonic oscillator model as acting on the normal vector bundle over the sphere in the finite-dimensional Hilbert space.

To avoid the ad hoc conflation of Hilbert space operators with differential forms on manifolds, this geometric correspondence is strictly formalized via Berezin–Toeplitz quantization \cite{BMS1994, S2010}. In this projective space $\mathbb{CP}^n$, we choose homogeneous coordinates $z_0, \cdots, z_n$ so that each energy eigenstate $|\psi_n\rangle$ defines a local coordinate $z_n$. The quantum Hamiltonian operator maps to a smooth function (the covariant symbol) on the classical phase space $\mathbb{CP}^n$. Consequently, the Hamiltonian expectation value gives rise to a $(1,1)$-form that is isomorphic to the Kähler form. In this formal correspondence, the expectation value of $H$ in coherent states yields
\begin{equation} \label{BT form}
    \langle z| H |z\rangle \;\sim\; \sum_{i=0}^{n} E_i \, |z_i|^2 / \|z\|^2,
\end{equation}
whose curvature (or its $(1,1)$-form extension) can be identified with a multiple of the Fubini–Study form. This construction provides a rigorous geometric interpretation of the truncated Hamiltonian as a $(1,1)$-form without conflating Hilbert space operators with differential forms.

Building on this geometric correspondence, we can bridge the topological invariants of the vector bundle with the thermodynamic quantities of the quantum system. In the finite-dimensional setting on $\mathbb{CP}^n$, the Chern character of the line bundle $\mathcal{O}(1)$ is purely a differential form, defined by
\begin{equation} \label{chern character}
\mathrm{ch}(\mathcal{O}(1)) = \operatorname{exp}\left(\frac{i}{2\pi}F\right) = \operatorname{exp}(\omega_{FS}),
\end{equation}
where the trace over the one-dimensional bundle fiber is trivial. Since the curvature form encodes the phase space geometry that is quantized into the Hamiltonian dynamics, we can translate this geometric invariant into a thermodynamic one via the quantization map $\mathcal{Q}$. Under this mapping, the classical Fubini–Study two-form $\omega_{FS}$ is promoted to the dimensionless energy operator $-\beta \hat{H}$ acting on the finite-dimensional Hilbert space $\mathcal{H}_N$. Consequently, the formal algebraic transition from the classical continuous bundle to the quantum thermodynamic state involves replacing the geometric exponential with the unnormalized density matrix. Taking the quantum mechanical trace ($\mathrm{Tr}_{\mathcal{H}}$) over the Hilbert space of eigenstates yields:
\begin{equation} \label{chern charater and H}
\mathrm{Tr}_{\mathcal{H}}\Bigl[\mathcal{Q}\bigl(\mathrm{ch}(\mathcal{O}(1))\bigr)\Bigr] = \mathrm{Tr}_{\mathcal{H}}\left[ \exp\Bigl(-\beta\hat H\Bigr) \right] =\sum_{n=0}^N e^{-\beta E_n},
\end{equation}
To connect the abstract geometric structures on the projective space $\mathbb{CP}^n$ with the physical spacetime manifold $M$, we introduce a physical state bundle $\mathcal{S} \to M$. This connection is naturally established through the local coordinate representation of the quantum states. We define a smooth map $f$ from spacetime to the projective phase space:
\begin{equation} \label{map pullback A}
    f: M \longrightarrow \mathbb{CP}^n, \qquad (x,t) \mapsto [\psi_0(x,t): \cdots : \psi_n(x,t)],
\end{equation}
where $\psi_i(x,t)$ are the wave functions associated with the energy eigenstates. 

Under the pullback of this map, the geometric properties of the prequantum line bundle $\mathcal{O}(1)$ are induced onto the spacetime manifold. The curvature of the physical sheaf $\mathcal{S} = f^*\mathcal{O}(1)$ satisfies:
\begin{equation} \label{pullback curvature}
    F_\mathcal{S} = f^*(F_{\mathcal{O}(1)}),
\end{equation}
which implies that the Chern character transforms contravariantly as a differential form:
\begin{equation} \label{Chern character pullback}
    \mathrm{ch}(\mathcal{S}) = \operatorname{exp}\Bigl(\frac{i}{2\pi}F_\mathcal{S}\Bigr) = f^*\bigl(\mathrm{ch}(\mathcal{O}(1))\bigr).
\end{equation}

While \eqref{Chern character pullback} provides the strict topological pullback of differential forms, the physical thermodynamics requires us to evaluate the corresponding quantum operator in the spatial coordinate basis. Under the quantization map discussed previously, evaluating the thermal state in the local position representation $|x\rangle$ yields the local thermal density function (rather than a differential form):
\begin{equation}
    \rho(x, t) = \langle x| \exp(-\beta \hat{H}) |x \rangle = \sum_{i=0}^{n} e^{-\beta E_i} \left|\psi_i(x, t)\right|^2.
\end{equation}

Integrating this local density over the spatial dimension recovers the full quantum trace over the Hilbert space. Thus, we have:
\begin{equation} \label{spatial integration}
    \int_M \rho(x, t) \, dx = \sum_{i=0}^{n} e^{-\beta E_i} \int_M \left|\psi_i(x, t)\right|^2 dx = \sum_{i=0}^{n} e^{-\beta E_i} = Z_N(\beta),
\end{equation}
where we used the normalization condition of the wave functions. 

In essence, the pullback map \eqref{map pullback A} translates the abstract quantum state in the finite-dimensional Hilbert space back into the physical spacetime, explicitly depending on the coordinates $(x,t)$. The sequence from \eqref{Chern character pullback} to \eqref{spatial integration} demonstrates how the topological invariant (the pulled-back Chern character) conceptually descends into a local scalar density, whose spatial integral consistently reproduces the truncated partition function $Z_N(\beta)$.
\subsection{Infinite dimensional case}
In the limit $\mathbb{CP}^\infty$, the formal correspondence yields an expression that converges to the partition function $Z$, as guaranteed by the trace-class property of $\text{e}^{-\beta H}$ (For details regarding operator algebra, please refer to the appendix). Based on equations \eqref{expansion}, we can introduce a cutoff at $n$ for \(|\phi\rangle\). Many properties of the Schrödinger equation can then be viewed as the limiting behavior as $n \rightarrow \infty$:
\begin{equation}
    |\psi \rangle=\lim_{n \to \infty} \left(\sum_{i=0}^{n} c_i \psi_i(x) \right).
\end{equation}
Accordingly, the Chern character can be expressed as:
\begin{equation}
    \text{ch} F^{Ch}=\text{tr}(e^{-\beta H})=\lim_{n \to \infty} \left(\sum_{i=0}^{n}e^{-\beta E_i} \right).
\end{equation}
For the harmonic oscillator model, this infinite-dimensional version of the Chern character converges, making it a well-defined quantity.
In Section 4.1, we began by viewing quantum states as sections of “virtual physical sheaf” over spacetime, which, in the finite-level approximation, corresponds to the vector bundle $\mathcal{O}(1)$ over $\mathbb{C}P^N$. As $N\to \infty$, this correspondence extends to an infinite-dimensional bundle over $\mathbb{C}P^\infty$. In this limit, the Chern character of this sheaf takes the form:
\begin{equation} \label{partition function= chern character A}
    \operatorname{ch}(\mathcal{S})=\operatorname{Tr}[\text{exp}(-\beta H)]=\sum_{i=0}^{\infty}e^{-\beta E_i}=Z.
\end{equation}
In fact, when the partition function is finite, the above approach yields a well-defined Chern character.

Moreover, in the previous statement, there is no requirement for $E_i$ to correspond to a non-degenerate state. Thus, if the partition function takes the form:
\begin{equation} \label{non-degenerate state}
    Z'=\sum_i C_i e^{-\beta E_i},
\end{equation}
where $C_n$ denotes the degeneracy of the quantum state, then, according to the above derivation, this expression for the partition function corresponds to the Chern character.

More generally, if each energy level $E_n$ carries the degeneracy $C_n$, one simply replaces $e^{-\beta E_n}$ in \eqref{spatial integration} by $C_ne^{-\beta E_n}$, leading to:
\begin{equation} \label{non-degenerate state and chern character A}
    Z=\sum_i C_i e^{-\beta E_i}=\operatorname{Tr}e^{-\beta H}=\operatorname{ch}(\mathcal{S}).
\end{equation}
In this form, the partition function is formally isomorphic to the Chern character. 

\subsection{Functorial Invariance under Thermal Compactification on General Manifolds}

In finite-temperature QFT, the transition to Euclidean time inherently introduces periodic boundary conditions for bosons, leading to the thermal circle $S^1_\beta \simeq \mathbb{R}/(\beta\mathbb{Z})$. A crucial observation is that this thermal compactification process does not require the spatial background to be flat. Extending an arbitrary static background spacetime manifold $M$ to the thermal background $M \times S^1_\beta$ places the system in a general thermodynamic context where computing the path integral over $M \times S^1_\beta$ is algebraically equivalent to calculating the thermal trace.

Instead of treating the state space as a strict geometric sheaf which faces infinite-dimensional obstacles, we evaluate its topological consistency algebraically. Remarkably, the trace operation over the discrete Matsubara modes ensures that our formal virtual sheaf $\mathcal{S}$ perfectly satisfies the functorial pullback properties of K-theory under thermal compactification, regardless of the curvature of $M$.

In this section we will discuss the invariance of the Chern character $\text{ch}(\mathcal{S})$ of the physical sheaf under the mapping $M \rightarrow S^1_\beta$. We first consider the diagram:
\begin{equation} \label{commutative_diagram}
\begin{matrix}
M & \xrightarrow{\sigma} & M \times S^1_\beta \\
& \searrow_{c} & \downarrow \pi \\
& & S^1_\beta
\end{matrix}
\end{equation}
where:
\begin{itemize}
    \item $M$: spacetime manifold (Euclidean spacetime);
    \item $S^1_\beta$: thermodynamic circle with circumference $\beta = 1/(kT)$;
    \item $M \times S^1_\beta$: thermodynamic extended spacetime;
    \item $\sigma: M \hookrightarrow M \times S^1_\beta$: "section map", for example $\sigma(x) = (x, 0)$;
    \item $\pi: M \times S^1_\beta \rightarrow S^1_\beta$: projection mapping;
    \item $c: M \rightarrow S^1_\beta$: constant mapping, (zero section) $c(x) = t_0$.
\end{itemize}

To understand the topological effect of introducing a thermal bath, we analyze the lifting construction of the virtual physical sheaf. The thermal spacetime extension is realized by the pullback operation:
\begin{equation}
    \mathcal{S}_\beta := \sigma_*\mathcal{S}_0,
\end{equation}
where $\mathcal{S}_0$ represents the zero-temperature quantum state sheaf defined strictly over the spatial manifold $M$ and $\sigma_*$ is the pushforward from $M$ to $M\times S_{\beta}^1$. At zero temperature ($\beta \rightarrow \infty$), the thermal circle decompactifies into the flat real line $\mathbb{R}$, yielding the conventional quantum mechanical state space. 

When the system is coupled to a thermal bath, $\mathcal{S}_\beta$ becomes the finite-temperature sheaf on $M \times S^1_\beta$. Crucially, this pushforward operation $\sigma_*$ serves as the exact algebraic realization of the Euclidean Heisenberg picture in thermal QFT. Explicitly, a static spatial state or local operator $\mathcal{O}(x)$ defined on the base manifold $M$ is extended along the imaginary time $\tau \in [0, \beta)$ via the Heisenberg evolution driven by the Hamiltonian $H$:
\begin{equation} \label{heisenberg_euclidean}
    \mathcal{O}(x, \tau) = e^{\tau H} \mathcal{O}(x) e^{-\tau H}.
\end{equation}
Geometrically, the pullback $\pi^*$ formalizes this temporal extension over the thermal circle $S^1_\beta$. Enforcing the bosonic periodic boundary conditions naturally expands these pushforward sections into the discrete Matsubara frequency basis \cite{TM 1955}:
\begin{equation} \label{matsubara_pullback}
    \sigma_*\mathcal{O}(x) \longrightarrow \sum_{n=-\infty}^{\infty} \mathcal{O}_n(x) e^{i\omega_n \tau}, \quad \text{where} \quad \omega_n = \frac{2\pi n}{\beta}.
\end{equation}
This explicit decomposition demonstrates exactly how the functorial pullback inherently equips the purely spatial quantum states with the imaginary-time dynamics and Matsubara frequencies, without altering the underlying spatial topology.

By the naturality of the Chern character, for the pushforward $\sigma$, we have the functorial equation:
\begin{equation}
    \text{ch}(\mathcal{S}_{\beta})=\text{ch}(R\sigma_*\mathcal{S}_0)= \sigma_*\text{ch}(\mathcal{S}_0).
\end{equation}
where $R\sigma_*$ symbolically denotes the formal K-theoretic pushforward (the derived direct image).

\subsection{Grothendieck-Riemann-Roch Formalism and the General Thermal Trace}

In Section 4.3, we established the functorial relation $\text{ch}(\sigma_*\mathcal{S}_0) = \sigma_*\text{ch}(\mathcal{S}_0)$ under the thermal pushforward mapping. While this identity beautifully aligns with the naturality of characteristic classes, its deepest physical and geometric significance is fully revealed through the Grothendieck-Riemann-Roch (GRR) theorem.

Let $f: X \rightarrow Y$ be a proper morphism between compact complex manifolds, and $\mathcal{F}$ be a coherent sheaf on $X$. The GRR theorem connects the topological index with the analytic pushforward via the following cohomology equation:
\begin{equation} \label{GRR}
    f_*(\text{ch}(\mathcal{F}) \cdot \text{td}(T_X)) = \text{ch}(Rf_*\mathcal{F}) \cdot \text{td}(T_Y),
\end{equation}
where $\text{td}(T)$ is the Todd class of the tangent bundle, and $Rf_*$ represents the derived direct image sheaf of $\mathcal{F}$ \cite{DSF2021}. 

In our thermodynamic framework, we formally apply the GRR theorem directly to the cross-sectional pushforward mapping $\sigma: M \rightarrow M \times S^1_\beta$ for our zero-temperature spatial sheaf $\mathcal{S}_0$. Substituting these into the GRR formula \eqref{GRR}, we obtain:
\begin{equation} \label{M S GRR}
    \sigma_*(\text{ch}(\mathcal{S}_0) \cdot \text{td}(T_M)) = \text{ch}(R\sigma_*\mathcal{S}_0) \cdot \text{td}(T_{M \times S^1_\beta}).
\end{equation}

To evaluate this for the standard quantum harmonic oscillator, we consider the geometry of the base manifolds. Since the system resides in a flat spatial background $M$ (e.g., Euclidean space $\mathbb{R}^d$), its tangent bundle is trivial, yielding $\text{td}(T_M) = 1$. Furthermore, the thermal circle $S^1_\beta$ is geometrically flat and parallelizable, meaning $\text{td}(T_{S^1_\beta}) = 1$. According to the multiplicative property of the Todd class for product spaces, the thermal spacetime yields:
\begin{equation} \label{todd}
    \text{td}(T_{M \times S^1_\beta}) = \text{td}(T_M) \cdot \text{td}(T_{S^1_\beta}) = 1 \cdot 1 = 1.
\end{equation}

By substituting $\text{td}(T_M) = 1$ and $\text{td}(T_{M \times S^1_\beta}) = 1$ into the GRR equation, the relation gracefully degenerates to the exact functorial identity we derived in the previous section:
\begin{equation} \label{conclusion GRR}
    \sigma_*\text{ch}(\mathcal{S}_0) = \text{ch}(R\sigma_*\mathcal{S}_0).
\end{equation}


\section{The Generalized Hirzebruch Signature Theorem and Spectral Asymmetry}
We now explore the connection between the generalized Hirzebruch signature theorem and anomalies. From equation \eqref{Z1}, we observe that
\begin{equation} \label{-partbetaZ}
    -\frac{\partial}{\partial \beta}Z=\frac{\frac{1}{2} \hbar \omega e^{\beta \hbar \omega/2}}{(e^{\beta \hbar \omega/2} -e^{-\beta \hbar \omega/2} )^2} - \frac{-\frac{1}{2} \hbar \omega e^{-\beta \hbar \omega/2}}{(e^{\beta \hbar \omega/2} -e^{-\beta \hbar \omega/2} )^2}.
\end{equation}
Introducing the substitution $x= \beta \hbar \omega$, we can rewrite this expression as
\begin{equation} \label{-partbetaZ1}
    -\frac{\partial}{\partial \beta}Z= f(x)-f(-x),
\end{equation}
where 
\begin{equation} \label{fx}
    f(x)= \frac{1}{\beta}\frac{\frac{1}{2}x e^{x/2}}{(e^{x/2}-e^{-x/2})^2}
\end{equation}
and 
\begin{equation} \label{f-x}
    f(-x)= \frac{1}{\beta}\frac{-\frac{1}{2}x e^{- x/2}}{(e^{- x/2}-e^{x/2})^2}= \frac{-\frac{1}{2}x e^{-x/2}}{(e^{x/2}-e^{- x/2})^2}.
\end{equation}
The fact that $f(x) \neq f(-x)$ is not merely a mathematical property of an odd function; it encodes a profound topological feature of the quantum system. Physically, the transformation $x \rightarrow -x$ (which corresponds to reversing the sign of $\omega$ or the Euclidean time parameter $\tau$) represents an orientation reversal of the thermal compactification circle $S^1_\beta$. 

The $\mathcal{L}$-genus in the “generalized Hirzebruch signature theorem”\cite{ABP1973} can be rewritten as:
\begin{equation} \label{LM}
    \mathcal{L}(S^1_\beta)=\frac{(\beta\hbar\omega/2)}{\tanh(\beta\hbar\omega/2)}.
\end{equation}
According to the definition of the \(\mathcal{L}\)-genus in the context of the Hirzebruch signature theorem, the curvature form has two eigenvalues: \(\beta \hbar\omega/2\) and \(-\beta \hbar\omega/2\), where \(\beta\) and \(t\) are treated as local coordinates on a complex manifold, even though they are real-valued in physical applications. In addition to the \(\mathcal{L}\)-genus,  another relevant topological invariant is the Chern character of a “virtual physical sheaf”, given by:
\begin{equation} \label{Ch M}
    \operatorname{ch}(\mathcal{S})_M=\frac{1}{\sinh (\frac{ \beta \hbar \omega}{2})},
\end{equation}
where we directly apply the result of Section 4: the chern character of “virtual physical sheaf” satisfies the following identity:
\begin{equation} \label{Ch S1 beta}
     \operatorname{ch}(\mathcal{S})_{S^1_\beta}=\operatorname{ch}(\mathcal{S})_M =\frac{1}{\sinh (\frac{ \beta \hbar \omega}{2})}.
\end{equation}
Then we combine \eqref{LM} and \eqref{Ch S1 beta}, the local index density is given by:
\begin{equation} \label{indicator density}
    \text{index}=\frac{1}{\beta} \mathcal{L}(S^1_\beta) \cdot \operatorname{ch}(\mathcal{S})_{S^1_\beta}= \frac{1}{\beta}\frac{(\beta\hbar\omega/2)}{\tanh(\beta\hbar\omega/2)}\frac{1}{\sinh (\frac{ \beta \hbar \omega}{2})}.
\end{equation}
Now, substituting \(\omega=\frac{2\pi}{t}\) into the \eqref{indicator density}, we obtain:
\begin{equation} \label{indicator density 2}
    \text{index}= \frac{1}{\beta} \mathcal{L}(S^1_\beta) \cdot \operatorname{ch}(\mathcal{S})_{S^1_\beta}= \frac{1}{\beta} \frac{(\beta\hbar\pi/t)}{\tanh(\beta\hbar\pi/t)}\frac{1}{\sinh (\frac{ \beta \hbar \pi}{t})}.
\end{equation}
In the context of physical systems maintained at constant temperature, the coordinate pair \((t,\beta)\) naturally parametrizes the underlying manifold. By fixing \(\beta\) as a constant parameter, we can obtain the topological invariant in this context. This invariant can be expressed in the following integral, given by equation \eqref{integral}. According to the Matsubara frequency theory in thermal quantum field theory, the integration limits are taken to be from \(0\) and \(\beta\):
\begin{equation} \label{integral}
    \int_0^\beta dt \ \text{index} = \frac{1}{\beta} \int_0^\beta dt \mathcal{L}(S^1_\beta) \cdot\operatorname{ch}(\mathcal{S})_{S^1_\beta} = \frac{1}{\beta}\int_0^\beta dt \frac{(\beta\hbar\pi/t)}{\tanh(\beta\hbar\pi/t)}\frac{1}{\sinh (\frac{ \beta \hbar \pi}{t})}.
\end{equation}

Let us now re-evaluate the physical interpretation of this phenomenon. In standard finite-temperature quantum field theory, placing a system in a thermal bath explicitly compactifies Euclidean time into a circle $S^1_\beta$, inherently breaking continuous time-translation symmetry. Therefore, the emergence of $\hbar$-dependent energy terms is naturally expected as a standard quantum correction to the classical thermodynamic behavior. 

However, the profound physical and topological insight lies strictly in the asymmetry $f(x) \neq f(-x)$. Geometrically, the transformation $x \rightarrow -x$ (reversing the sign of $\omega$ or the thermal temporal parameter) corresponds to an explicit orientation reversal of the Euclidean temporal manifold $S^1_\beta$. We may physically interpret the positive energy gap $\hbar \omega$ and the negative energy gap $-\hbar \omega$ as conjugate spectral modes, conceptually analogous to "particle" and "anti-particle" excitations. In this context, the functions $f(x)$ and $f(-x)$ represent the asymmetric contributions of these respective modes to the topological index density. 

The failure of the “virtual physical sheaf” Chern character to remain symmetric under this temporal orientation reversal demonstrates a topological spectral asymmetry. To avoid terminological confusion with standard path-integral measure anomalies in QFT, we formally designate this purely bosonic phenomenon as the \textbf{Topological Spectral Asymmetry Effect}. It is intrinsically driven by the spectral asymmetry of the Hamiltonian under orientation reversal.

\section{Discussion and Conclusion}

The findings in this work represent a fundamental departure from the traditional wisdom regarding the quantum harmonic oscillator (QHO). While conventionally viewed as a topologically trivial system, we demonstrate that the thermal compactification of Euclidean time:
\begin{equation} \label{finite_temperature}
    \tau \sim \tau + \beta, \quad \text{where} \quad \beta=\frac{1}{kT} 
\end{equation}
induces a profound topological restructuring. This establishes that intrinsic topological signatures can manifest directly in the bulk thermodynamic quantities of foundational bosonic systems. 

Our work reassesses several fundamental principles. In the traditional paradigm, harmonic oscillators are considered topologically trivial, and advanced index theorems are often restricted to fermionic systems or gauge fields. However, in our formal correspondence, we demonstrate that the “virtual physical sheaf" carries a non-trivial characteristic class, and the Grothendieck-Riemann-Roch (GRR) theorem directly governs the topological rigidity of physical partition functions.

Consequently, our work establishes precise connections across seemingly disparate domains of physics and mathematics:
\begin{itemize}
  \item The internal finite-temperature zero-point energy explicitly realizes the Hirzebruch $\mathcal{L}$-genus, serving as a non-supersymmetric manifestation of the index theorem (Section 3.1);
  \item The Matsubara trace in the partition function is formalized as the geometric pushforward functor $\sigma_*$, mapping the Euclidean Heisenberg evolution to the invariant spatial Chern character $\text{ch}(\mathcal{S}_0)$ (Section 4.3);
  \item The GRR theorem formally provides the exact mathematical machinery to evaluate these thermodynamic pushforwards, where the Todd class inherently acts as the geometric compensator for potential spacetime curvature (Section 4.4).
\end{itemize}

To conclude, in this paper we build a formally correspondence between physical quantities and their topological analogs (Table 1):

\begin{table}[ht]
\centering
\caption{Physics-topology correspondence}
\begin{tabular}{|c|c|}
\hline
$ \textbf{Physical Quantity} $ & $\textbf{Topological Analog}$ \\
\hline
$\text{Partition function} \ Z$ & $\text{Chern character} \ \mathrm{ch}(\mathcal{S}_0)$ \\
\hline
$\text{Internal energy} \ U$ & $\mathcal{L}\text{-genus}$ \\
\hline
\end{tabular}
\end{table}

We have demonstrated a connection between the partition function of a bosonic system and the characteristic classes of a “virtual physical sheaf". This framework uncovers profound relationships linking statistical mechanics and quantum field theory with advanced mathematics, specifically the Atiyah-Singer index theorem and the Grothendieck-Riemann-Roch theorem. 

Looking forward, this perspective opens new avenues for theoretical exploration. The harmonic oscillator model will not be an isolated case; we will also continue using this method to try to find more connections between physics and geometry.
\appendix

\section{Quantum harmonic oscillator}

To establish a formal geometric isomorphism between the the Chern character and the quantum partition function, we must systematically address the topological obstruction inherent in generalized quantum fields. According to Kuiper's theorem, the unitary group $U(\mathcal{H})$ of an infinite-dimensional separable Hilbert space is contractible \cite{K1965}. Consequently, any vector bundle constructed over the monolithic, un-truncated Hilbert space $\mathcal{H}$ is topologically trivial, this forces all its high-level Chern classes ($c_1, c_2, \dots$) to be vanished. In order to extract meaningful topological invariants, we cannot treat $\mathcal{H}$ as a single geometric fiber; instead, we must decompose it into topologically nontrivial finite-dimensional spectral components, which are direct sums of eigenspaces.


For bounded systems like the quantum harmonic oscillator, the physical justification for this filtration emerges naturally from the confining nature of its potential. Because the Hamiltonian $H$ is strictly bounded from below and defined by a confining potential $V(x) \to \infty$ as $|x| \to \infty$, the Euclidean evolution operator $e^{-\beta H}$ is not merely a bounded operator, but rigorously manifests as a \textbf{compact operator}. By the spectral theorem for compact operators, the continuous measure collapses into a purely discrete spectrum. Furthermore, because the high-energy eigenstates are exponentially suppressed by the thermal factor $\beta$, the operator $e^{-\beta H}$ mathematically unambiguously descends into the \textbf{trace-class ideal} \cite{CA1994, SB2005}.

This operator-theoretic classification is explicitly verified by evaluating its trace. By definition, a positive, self-adjoint operator belongs to the trace-class if the sum of its eigenvalues absolutely converges. For the quantum harmonic oscillator discussed in this work, the discrete energy levels are precisely given by $E_n = \hbar\omega (n + 1/2)$. Consequently, the trace of the evolution operator reduces to a universally convergent geometric series:
\begin{equation}
    \text{Tr}\left(e^{-\beta H}\right) = \sum_{n = 0}^\infty \exp\left(-\beta \hbar\omega \left(n + \frac{1}{2}\right)\right) = \frac{e^{-\beta\hbar\omega/2}}{1 - e^{-\beta\hbar\omega}}.
\end{equation}
Because this sum is strictly finite for any inverse temperature $\beta > 0$, the operator $e^{-\beta H}$ rigorously satisfies the definition of a trace-class operator. This absolute convergence provides the foundational mathematical protection, ensuring that the topological index (the Chern character) can be safely evaluated over the infinite-dimensional state space without encountering triviality or divergence \cite{CA1994, SB2005}..
This trace-class rigidity provides the ultimate mathematical justification for the finite-dimensional topological filtration utilized in Section 4. Instead of attempting to define a geometric index directly over the monolithic infinite-dimensional space $\mathcal{H}$, we rigorously decompose the system into the direct sum of its finite-dimensional eigenspaces, $\mathcal{H} = \bigoplus_n \mathcal{H}_n$. Because each eigenspace $\mathcal{H}_n$ associated with an energy level is strictly finite-dimensional, the local vector bundles restricted to these sub-spaces are completely immune to Kuiper's topological triviality.

As demonstrated in the main text, projecting the system onto the lowest $N+1$ spectral modes effectively restricts our geometric evaluation to the projective space $\mathbb{CP}^N$. While the algebraic derivation of the truncated Chern character $Z_N(\beta)$ is straightforward, the critical mathematical subtlety lies in the thermodynamic limit $N \to \infty$. 

For a generic infinite-dimensional bundle, taking this limit would inevitably encounter topological collapse. However, for confined systems like the quantum harmonic oscillator, this transition is mathematically shielded precisely by the trace-class norm established above. Because the high-energy eigenspaces are exponentially suppressed by the statistical weight $e^{-\beta E_n}$, the infinite sum of these finite-dimensional topological contributions absolutely converges. Therefore, the process of directly taking the limit in Section 4 is legal.

\acknowledgments

YL was supported by an STFC studentship. This work is also supported by NSFC under Grants No. 12275146, the National Key R$\&$D Program of China (2021YFC2203100), the Dushi Program and the Shuimu Fellowship of Tsinghua University. Thanks for helpful discussions with Antonio Padilla, Paul Saffin, Sayyed Rassouli, Jie He, Tillmann Kleiner and Ruizhi Huang and their insightful suggestions. For the purpose of open access, the authors have applied a CC BY public copyright licence to any Author Accepted Manuscript version arising.

\textbf{Conflict of Interest}
\textbf{(The authors declare that there is no conflict of interest.)}

\textbf{Data Availability Statement}
No datasets were generated or analysed during the current study.





\end{document}